\documentclass[fleqn,twoside]{article}
\usepackage{espcrc2}

\usepackage{graphicx}
\usepackage[figuresright]{rotating}

\title{Advances in Open Charm Physics at CLEO-c}

\author{Paras Naik\address[BRI]{University of Bristol, Bristol BS8 1TL, United Kingdom}\\}
       
\begin{document}

\begin{abstract}
We present a survey of CLEO-c open charm results. The unprecedented full data sample includes over $10$ million $D$ mesons and approximately $1.2$ million $D_s$ mesons. These results substantially extend the reach and understanding of heavy flavor physics. The world community will benefit as results from CLEO-c extend the reach of the Belle experiment at KEK and the LHCb experiment at CERN and lay foundations for the physics program of the BESIII experiment in China.
\vspace{1pc}
\end{abstract}

\maketitle

\section{Introduction}

The CLEO-c experiment \cite{CLEO} at the {CESR-c} (Cornell Electron Storage Ring - charm) $e^+e^-$ collider has collected large data samples
in the charm threshold energy region. Open charm samples have allowed the CLEO-c experiment to have an impact on measurements of the Cabibbo-Kobayashi-Maskawa (CKM) matrix elements, improved our knowledge of hadronic $D$ and $D_s$ decays, and enabled studies of rare charm processes.

Precision measurements of the weak phases that compose the unitarity triangle ($\alpha$, $\beta$, and $\gamma$) 
allow us to test the internal consistency of the CKM model and search for 
signatures of New Physics.  
The CKM phase $\gamma$ is only constrained by direct measurements to $({71 ^ { + 21} _ {- 25} })^{\circ}$ \cite{CKMFitter}. 
The most promising methods of determining the CKM phase
 $\gamma$ exploit the interference within $ B^{\mp}\to 
DK^{\mp}$ decays, where the neutral $D$ meson is a $ D^{0}$ or $ \bar{D}^{0}$. The most straightforward of these
strategies considers two-body final states of the $D$ meson, but additional information can be gained from 
strategies that consider multi-body final states.
The parameters 
associated with the specific 
final states needed for these analyses can be extracted from correlations within 
CLEO-c  $\psi(3770)$ data. 

Lattice QCD theory offers the prospect of a systematically-improvable method of calculating
hadronic properties from first principles. In the past decade, theoretical and technological
advances have allowed Lattice QCD to provide precise predictions that estimate systematic uncertainties reliably. 
An exciting application of these results allows the extraction of electroweak parameters in the $B$ meson system. However, it is desirable to
test them elsewhere, for example in the charm system, and CLEO-c provides excellent data for these tests.

\section{Detector and Data Samples}
The CLEO-c detector was a symmetric general purpose detector that provided 
excellent electromagnetic calorimetry, charged particle tracking and identification, and near 4$\pi$ solid angular 
coverage. 
The experiment is described in detail elsewhere \cite{CLEO}. 

The relevant data sets for the following analyses
were collected at center of mass energies of approximately $3.77$ GeV (the peak of the $\psi(3770)$
resonance) and $4.17$ GeV. The former data set is used for $D^0$ and  $D^+$ analyses, and the latter
for  $D_s$ physics. Except where noted, all analyses use the full CLEO-c data sets of $818~\rm{pb}^{-1}$ at $3.77$ GeV
and $600~\rm{pb}^{-1}$ at $4.17$ GeV. 

At 3.77 GeV the only allowed open charm final states are $D^0{\bar{D}^0}$ and $D^+{D^-}$. At 4.17 GeV the only allowed states involving a $D_s$ meson are $D_s^+{D_s^-}$ and $D_s^\pm{D_s^{*\mp}}$. 
Since only two $D$ mesons of opposite flavor are produced, we may use the double tagging technique pioneered by Mark III \cite{2}.
This allows CLEO-c to obtain very clean samples of charm meson decays, with minimal combinatoric background.
In particular, using neutral $D$ meson tags allow us to exploit quantum correlations of the
initial state ($\psi(3770) \to D^0{\bar{D}^0}$).
Also, neutrinos and $K_L$ mesons may be inferred via the full reconstruction
of the observed particles in an event.

\section{Strong Phases for $\gamma$/$\phi_3$ Measurements}

\subsection{$D \to K^-\pi^+$}

Atwood, Dunietz and Soni (ADS) \cite{ADS} have suggested considering $D$ decays to non-$CP$ eigenstates as a way of maximizing sensitivity 
to $\gamma$ measured via studies of the rates of $B^{\mp} \to D K^{\mp}$ decays. $D$ meson final states such as $K^{-}\pi^{+}$, which may arise from either a Cabibbo favored $D^{0}$ decay or a doubly 
Cabibbo suppressed ${\bar{D}^0}$ decay, can lead to large interference effects and hence provide particular sensitivity to 
$\gamma$. 
The interference phase, $\delta_D^{ K \pi}$, also relates the $D^0$ mixing
parameters $y$ and $y^\prime$. The differences in the effective branching fraction for $K^-\pi^+$ decay opposite
$CP$-even and -odd eigenstates, semileptonic decays,
 and $K^+\pi^-$ are sensitive to both $\delta_D^{ K \pi}$ and to $D^0$ mixing
parameters. With $281~\rm{pb}^{-1}$ of data, 
CLEO-c has measured $\delta_D^{ K \pi}$ to be $(22^{+14}_{-15})^{\circ}$ through a quantum correlated analysis of completely reconstructed $\psi(3770) \to D\bar{D}$ decays \cite{TQCA}.

\subsection{Multi-body Decays}

The ADS formalism can be extended by
considering multi-body decays of the $D$ meson. 
However, a multi-body $D$-decay amplitude is potentially 
different at any point within the decay phase space, because of the contribution of intermediate resonances. It is shown in 
Ref.~\cite{AS} how the rate equations for the two-body ADS method should be modified for use with multi-body final states. 
The result is an effective average phase and a ``coherence
factor'' which reflects the dilution of total interference relative to the expectation for a simple
two-body decay. 
CLEO-c has measured these for the $D^0 \to K^-\pi^+\pi^0$ and $D^0 \to K^-\pi^+\pi^+\pi^-$  decays,
observing significant coherence in the former \cite{16}.

\subsection{$K^0_{S,L} \pi^+ \pi^-$ and  $K^0_{S,L} K^+ K^-$ }
Dalitz plot analyses of the three-body decay ${D} \to K_S^0 \pi^+\pi^-$
together with studies of $B^{\mp}\to D K^{\mp}$ processes currently provide the best
measurements of the CKM weak phase $\gamma$
\cite{BelleA,BabarNewA}.
However, ${D} \to K_S^0 \pi^+\pi^-$  Dalitz plot analyses are sensitive to the choice of the
model used to describe the three-body decay, which currently introduces a model systematic
uncertainty on the determination of  $\gamma$ which is greater than $5^{\circ}$ \cite{BabarNewA}. 
For LHCb and future Super-$B$ factories, this
uncertainty will become a major limitation.
A model independent approach to understanding the $D$ decay has been proposed by
Giri and further
investigated by Bondar \cite{GB}, which takes advantage of the quantum correlated $D^0/\bar D^0$
CLEO-c data produced at the $\psi(3770)$ resonance.

\begin{figure*}[htb]
\center\includegraphics[scale=.35]{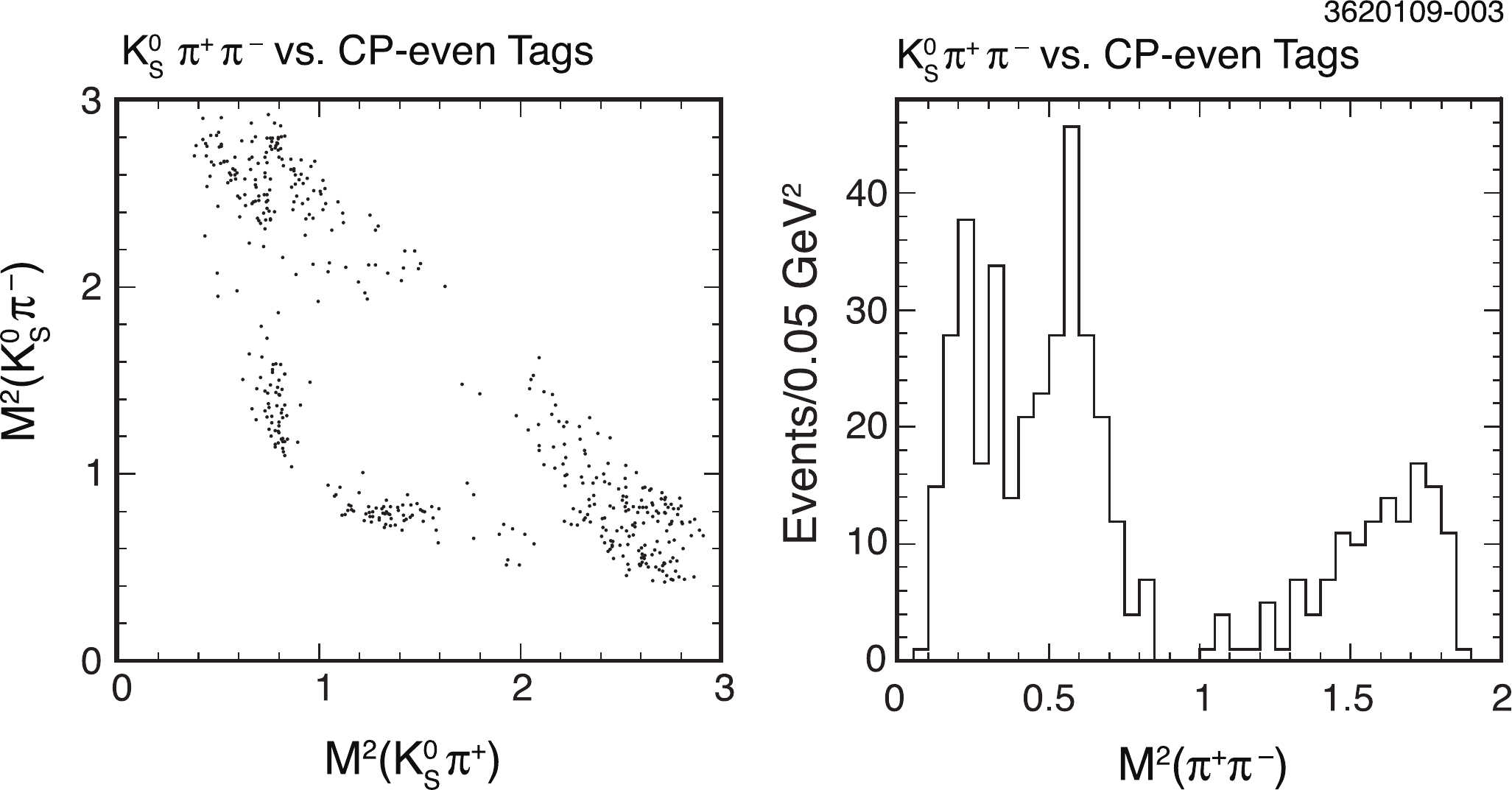}~~~~~\includegraphics[scale=.35]{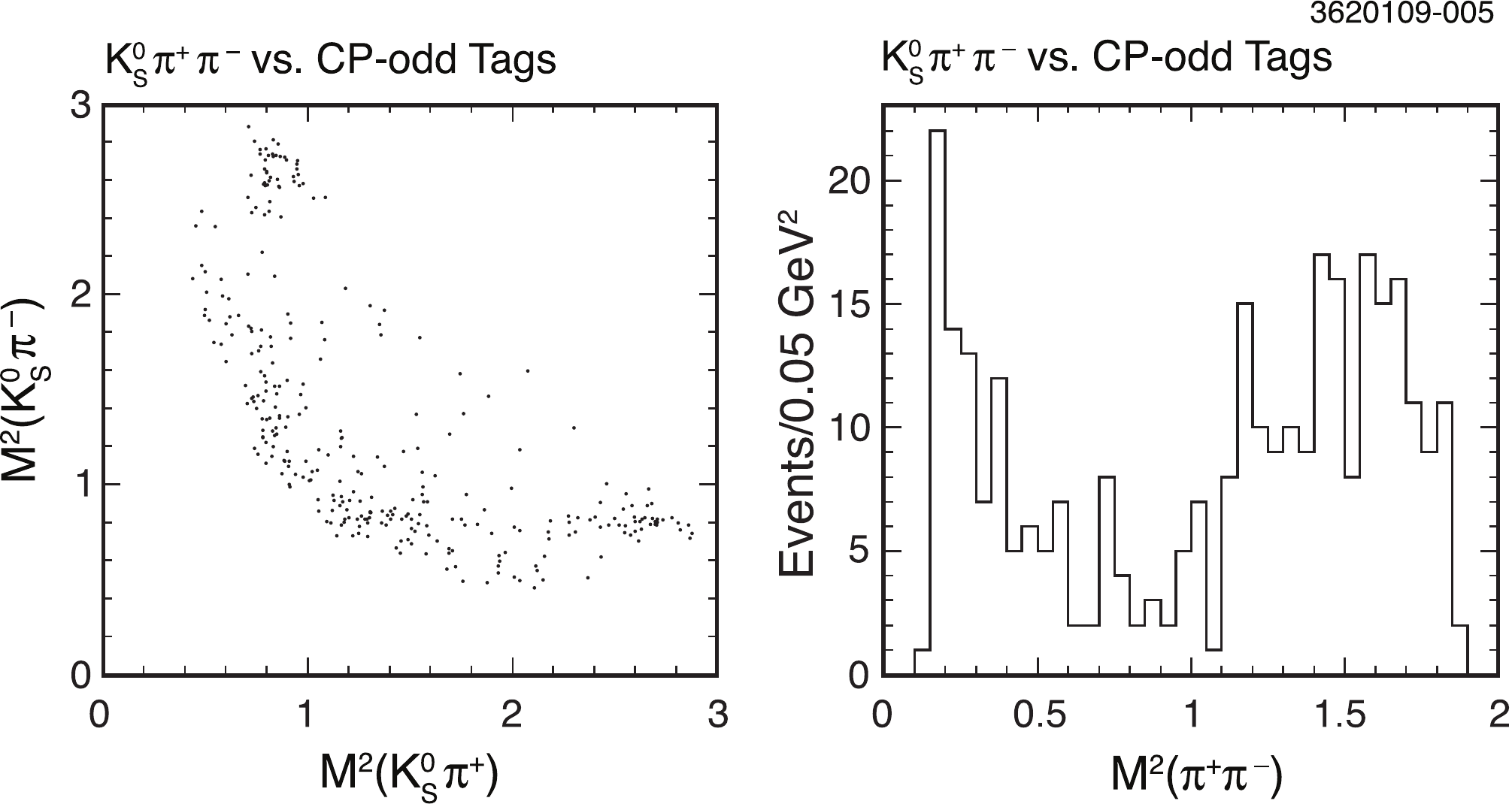}
\caption{Effects of $CP$ correlation on the Dalitz plot of the decay $D^0 \to K^0_S\pi^+\pi^-$. The $K^0_S\rho$ component, clearly
visible when $K^0_S\pi^+\pi^-$ recoils against a CP-even tag, disappears opposite a CP-odd tag.}
\label{fig3}
\end{figure*}

CLEO-c has performed
this measurement for the $K^0_{S,L} \pi^+ \pi^-$ decay \cite{17} and for the $K^0_{S,L} K^+ K^-$ mode \cite{NEWKsKK} which may also be used in this method.
Fig. \ref{fig3} shows the effect of the $CP$ correlations on the $D^0 \to K^0_S \pi^+ \pi^-$ Dalitz plot. Up to small
effects $K^0_L \pi^+ \pi^-$ has a $CP$ structure opposite that of $K^0_S \pi^+ \pi^-$, and similarly for $K^0_L K^+ K^-$ and $K^0_S K^+ K^-$.

\subsection{Impact on $\gamma$ Measurement}
The $D^0 \to K^0_S \pi^+ \pi^-$ analysis is expected to reduce the current 7-9$^\circ$
model uncertainties from BaBar and Belle measurements \cite{BelleA,BabarNewA} to around $2^\circ$ \cite{17,19} at LHCb. The $K^-\pi^+$ and
multi-body coherence factor measurements are projected to improve the precision of
LHCb in $B \to DK$ by 8-35\% (depending on unknown $B$ decay parameters) to 2.2-3.5$^\circ$ after  $10~\rm{fb}^{-1}$ of LHCb data taking \cite{20}.

\section{Leptonic Decays and Decay Constants}

The decay $X^+ \to \ell^+\nu$ of pseudoscalar $X$ involves a hadronic current (parametrized by the single
``decay constant'' $f_X$) and a well-understood Standard Model leptonic current.
The branching fraction can be written as:
\begin{equation}
\mathcal{B}(X^+ \to \ell^+\nu)=f_X^2 |V|^2 \frac{G_F^2}{8\pi}m_Xm_\ell^2(1-\frac{m_\ell^2}{m_X^2})^2~~
\end{equation}
where $V$ is the relevant element of the CKM matrix connecting the valence quarks of $X$ ($V_{cd}$ and
$V_{cs}$ for $D^+$ and $D^+_s$, respectively). 
The decay constant
is essentially a measure of the wave function of the meson at zero separation between the quarks. This
makes the decay constant relevant for processes where the relevant length scales are much smaller than the
hadron size, including the loop diagrams for $B^0_d$ and $B^0_s$ mixing (our primary source of
information on $V_{td}$).

We can determine $f_X^2|V|^2$ by measuring the branching fraction of $X$. Thus either the decay constant or the CKM element can be determined if we know the other quantity. CLEO-c has measured the  $D^+$ and $D^+_s$ decay constants in
multiple decay modes. 
The results of all CLEO-c leptonic branching fraction measurements and the corresponding
decay constants are shown in Table \ref{table:1a}. The values of input parameters used to obtain these values
are listed in the corresponding papers \cite{3,4,5,6}.

\section{Exclusive Semileptonic Decays}

\begin{figure*}[htb]
%\vspace{9pt}
\center\includegraphics[scale=.4]{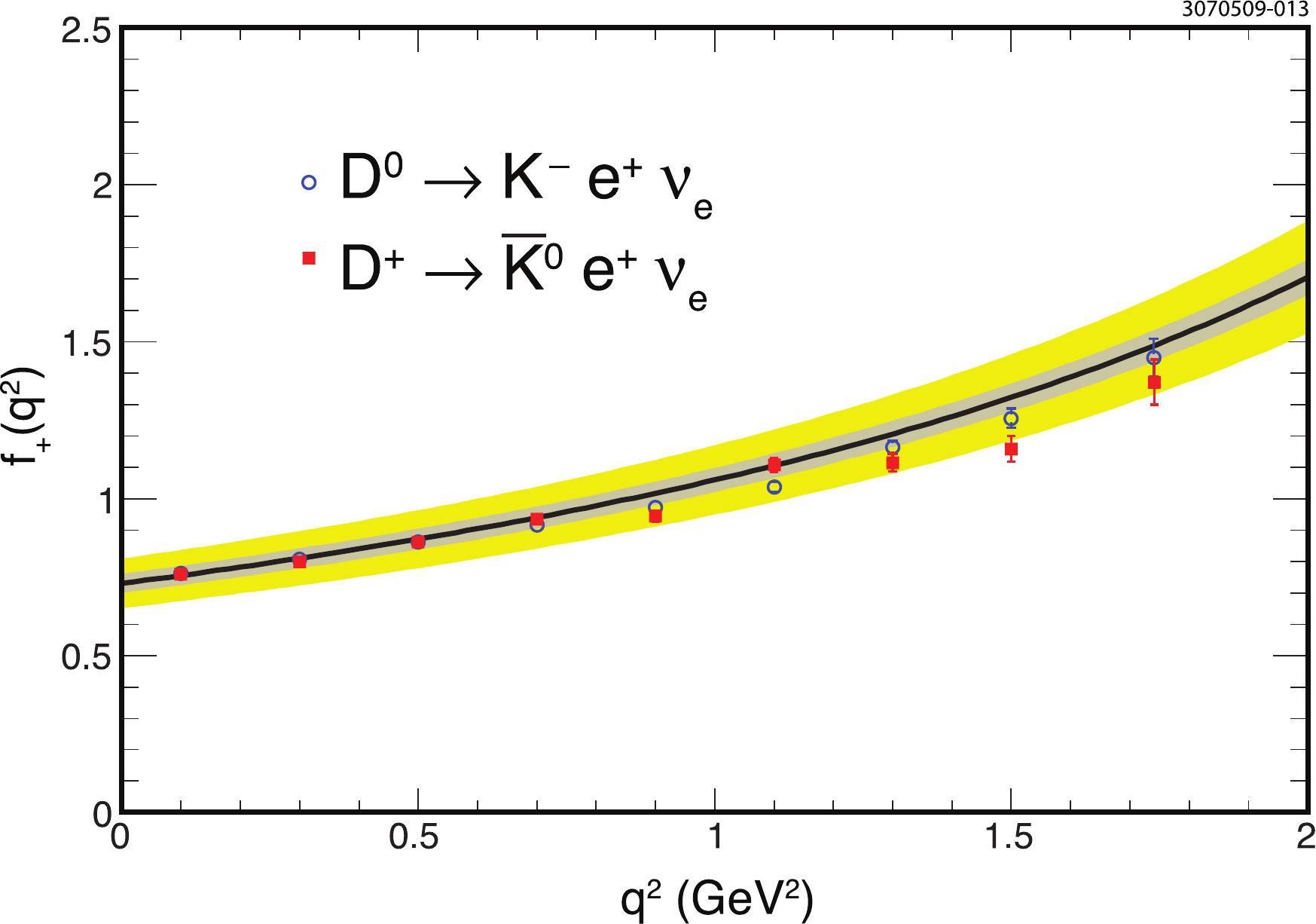}~~~~~\includegraphics[scale=.4]{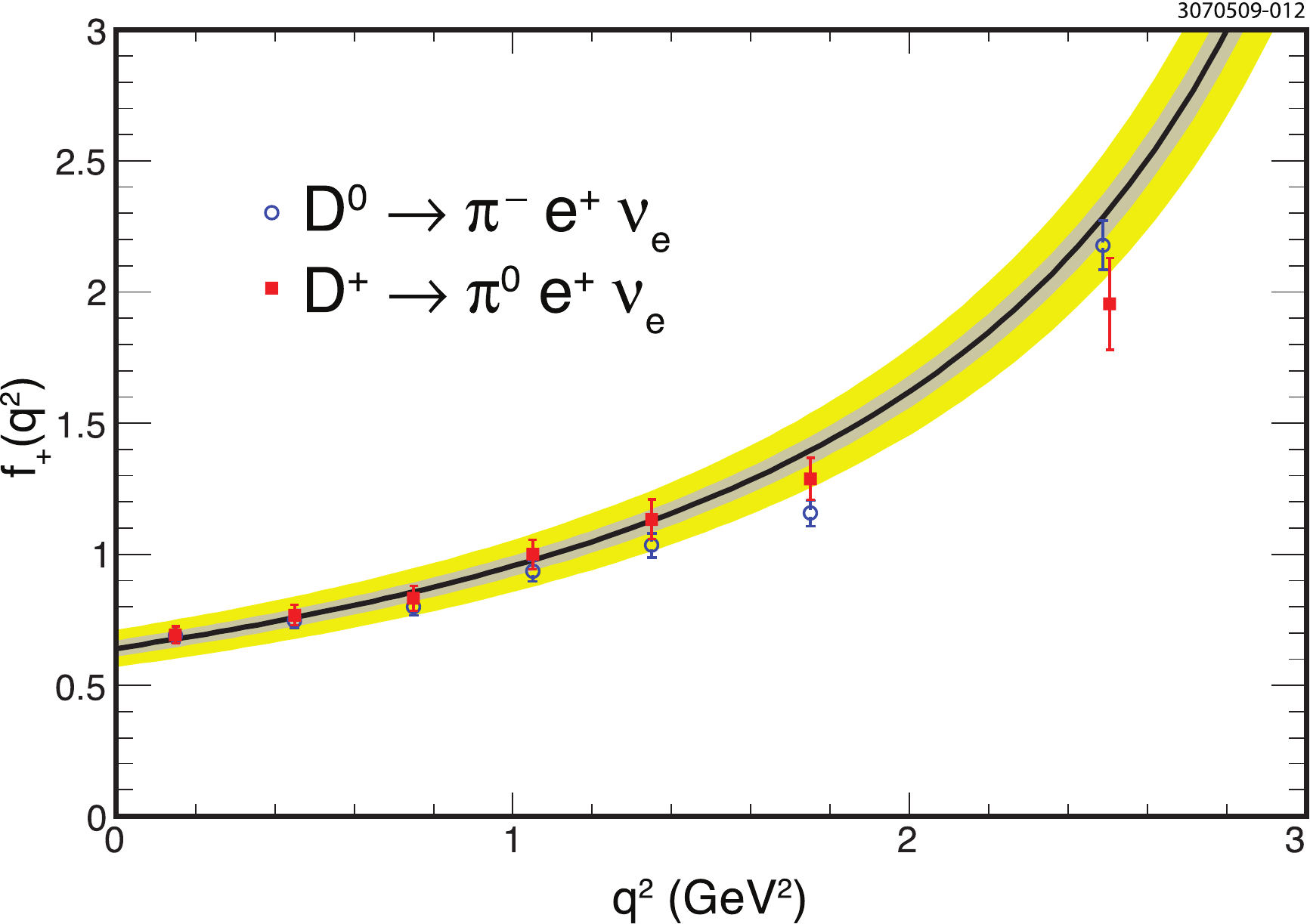}
\caption{$D$ semileptonic decay form factors for kaon (left) and pion (right) decays as a function of $q^2$. The points
are CLEO-c data, and the bands are from a lattice QCD prediction.}
\label{fig2}
\end{figure*}

Exclusive semileptonic decays are more difficult to parameterize than leptonic decays since there more than two particles in the final state. 
The partial width for the decay $X \to X^\prime\ell\nu$,
where $X$ and $X^\prime$ are pseudoscalars, can be written as:
\begin{equation} \label{eqSL}
\frac{d\Gamma(X \to X^\prime\ell\nu)}{dq^2}=\frac{G_F^2}{24\pi^3}[f_+^{X\to X^\prime}(q^2)|V|]^2p^3_X~~
\end{equation}
in the limit where the charged lepton mass is negligible. In Equation~\ref{eqSL}, $q^2$ is the invariant mass squared
of the $\ell\nu$ system, $|V|$ is the relevant CKM matrix element for the weak transition, and $f^{X \to X^\prime}_+$
is a form factor which represents the hadronic physics interactions. As in the leptonic decay case, input for either $|V|$ or $f_+$ allows determination of the other. 
We analyze $D \to (K, \pi)e^+\nu$ decays \cite{8} in order to determine the form factors in bins of $q^2$.
Reasonable agreement on the form factor
shape and normalization is found with a lattice QCD prediction from the FNAL, MILC, and
HPQCD collaborations \cite{11}. A comparison of our results to the theory can be seen in Figure \ref{fig2}. Using lattice predictions for
$|f_+(0)|$, values for $|V_{cd}|$ and $|V_{cs}|$ are also obtained, which are limited by lattice uncertainties.

\begin{table*}[htb]
\caption{CLEO-c measurements of $D^+$ and $D^+_s$ leptonic decay branching fractions and decay constants.}
\label{table:1a}
\newcommand{\m}{\hphantom{$-$}}
\newcommand{\cc}[1]{\multicolumn{1}{c}{#1}}
\renewcommand{\tabcolsep}{2pc} % enlarge column spacing
\begin{tabular}{@{}ll}
\hline
&CLEO-c Result\\
\hline
$B(D^+ \to \mu^+\nu)$ \cite{4} &$(3.82 \pm 0.32 \pm 0.09) \times 10^{-4} $\\%&\\
$B(D^+_s \to \mu^+\nu)$ \cite{5} &$ (5.65 \pm 0.45 \pm 0.17) \times 10^{-3}$\\%&\\
$B(D^+_s \to \tau^+\nu) ({\rm from}~\tau^+ \to \pi^+\bar{\nu})$ \cite{5}& $(6.42 \pm 0.81 \pm 0.18) \times 10^{-2}$\\%&\\
$B(D^+_s \to \tau^+\nu) ({\rm from}~\tau^+ \to e^+\nu\bar{\nu})$ \cite{6}& $(5.30 \pm 0.47 \pm 0.22)  \times 10^{-2}$\\%&\\
$B(D^+_s \to \tau^+\nu) ({\rm from}~\tau^+ \to \rho^+\bar{\nu})$ \cite{3} &$(5.52 \pm 0.57 \pm 0.21)  \times 10^{-2}$\\%&\\
\hline
$f_{D^+}$ &($205.8 \pm 8.5 \pm 2.5$) MeV \\ 
$f_{D^+_s}$ (combined) &($259.0 \pm 6.2 \pm 3.0$) MeV \\ 
\hline
\end{tabular}\\[2pt]
\end{table*}

\begin{thebibliography}{9}

\bibitem{CLEO} Y.~Kubota {\it et al.}, Nucl. Instrum. Meth. Phys. Res., Sect. A {\bf 320}, 66 (1992); \\D.~Peterson {\it et al.}, Nucl. 
Instrum. Meth. Phys. Res., Sect. A {\bf 478}, 142 (2002).%1
\bibitem{CKMFitter} J. Charles {\it et al.}, CKMfitter Group, Eur. Phys. J. \textbf{C41}, 1 (2005); \\Updated results and plots at {\tt 
http://ckmfitter.in2p3.fr}.%2
\bibitem{2} R. M. Baltrusaitis {\it et al.}, Phys. Rev. Lett. \textbf{56}, 2140 (1986); Adler J. {\it et al.}, Phys. Rev. Lett. \textbf{60}, 89 (1988).%3
\bibitem{ADS} D. Atwood, I. Dunietz and A. Soni, Phys. Rev. Lett. \textbf{78}, 3257 (1997).%4
\bibitem{TQCA} J.L. Rosner et al., Phys. Rev. Lett. \textbf{100}, 221801 (2008); D. Asner et al., Phys. Rev. D \textbf{78}, 012001 (2008).%5
\bibitem{AS} D. Atwood and A. Soni, Phys Rev. D \textbf{68}, 033003 (2003).%6
\bibitem{16} N. Lowrey {\it et al.}, Phys. Rev. D \textbf{80}, 031105 (2009).%7
\bibitem{BelleA} Belle Collaboration, arXiv:0803.3375v1 [hep-ex] (2008).%8
\bibitem{BabarNewA} B. Aubert {\it et al.}, Phys. Rev. D \textbf{78}, 034023 (2008).%9
\bibitem{GB} Giri {\it et al.}, Phys. Rev. D \textbf{68}, 054018 (2003); Bondar {\it et al.}, Eur. Phys. J. C \textbf{47}, 347-353 (2006).%10
\bibitem{17} R. A. Briere {\it et al.}, Phys. Rev. D \textbf{80}, 032002 (2009).%11
\bibitem{NEWKsKK} J. Libby {\it et al.}, arXiv:1010.2817 [hep-ex] (2010).%12
\bibitem{19} J. Libby, CERN-LHCb-2007-141, 2007, unpublished.%13
\bibitem{20} K. Akiba {\it et al.}, CERN-LHCb-2008-031, 2008, unpublished.%14
\bibitem{3} P. Naik {\it et al.}, Phys. Rev. D \textbf{80}, 112004 (2009). %15
\bibitem{4} B. I. Eisenstein {\it et al.}, Phys. Rev. D \textbf{78}, 052003 (2008). %16
\bibitem{5} J. P. Alexander {\it et al.}, Phys. Rev. D \textbf{79}, 052001 (2009). %17
\bibitem{6} P. U. E. Onyisi {\it et al.}, Phys. Rev. D \textbf{79}, 052002 (2009). %18
\bibitem{8} D. Besson {\it et al.}, Phys. Rev. D \textbf{80}, 032005 (2009). %19
\bibitem{11} C. Aubin {\it et al.}, Phys. Rev. Lett. \textbf{94}, 011601 (2005). %20
\end{thebibliography}
\end{document}